\documentclass[10pt,twocolumn,letterpaper]{article}

\usepackage{wacv}
\usepackage{times}
\usepackage{epsfig}
\usepackage{graphicx}
\usepackage{amsmath}
\usepackage{amssymb}

\usepackage{floatrow}
\usepackage{booktabs}
\usepackage[accsupp]{axessibility}  

\def\wacvPaperID{1356} 

\wacvfinalcopy 

\ifwacvfinal
\usepackage[breaklinks=true,bookmarks=false]{hyperref}
\else
\usepackage[pagebackref=true,breaklinks=true,colorlinks,bookmarks=false]{hyperref}
\fi

\ifwacvfinal
\fi

\begin{document}

\title{AFTer-UNet: Axial Fusion Transformer UNet for Medical Image Segmentation}

\author{
Xiangyi Yan \hspace{9pt} Hao Tang \hspace{9pt} Shanlin Sun \hspace{9pt} Haoyu Ma \hspace{9pt} Deying Kong \hspace{9pt} Xiaohui Xie\\
University of California, Irvine\\
{\tt\small \{xiangyy4, htang6, shanlins, haoyum3, deyingk, xhx\}@uci.edu}

}

\maketitle
\thispagestyle{empty}

\begin{abstract}
   
    Recent advances in transformer-based models have drawn attention to exploring these techniques in medical image segmentation, especially in conjunction with the U-Net model (or its variants), which has shown great success in medical image segmentation, under both 2D and 3D settings. Current 2D based methods either directly replace convolutional layers with pure transformers or consider a transformer as an additional intermediate encoder between the encoder and decoder of U-Net. However, these approaches only consider the attention encoding within one single slice and do not utilize the axial-axis information naturally provided by a 3D volume. In the 3D setting, convolution on volumetric data and transformers both consume large GPU memory. One has to either downsample the image or use cropped local patches to reduce GPU memory usage, which limits its performance. In this paper, we propose Axial Fusion Transformer UNet (AFTer-UNet), which takes both advantages of convolutional layers' capability of extracting detailed features and transformers' strength on long sequence modeling. It considers both intra-slice and inter-slice long-range cues to guide the segmentation. Meanwhile, it has fewer parameters and takes less GPU memory to train than the previous transformer-based models. Extensive experiments on three multi-organ segmentation datasets demonstrate that our method outperforms current state-of-the-art methods.
\end{abstract}

\vspace{-0.9em}
\section{Introduction}
Medical image segmentation is an essential procedure in many modern clinical workflows. It can be used in many applications, including diagnostic interventions, treatment planning and treatment delivery \cite{gibson2018automatic, tang2019automatic}. These image analyses are usually carried out by experience doctors. However, it is labor-intensive and time-consuming, since a 3D CT volume can contain up to hundreds of 2D slices. Therefore, developing robust and accurate image segmentation tools is a fundamental need in medical image analysis \cite{sun2020attentionanatomy, tang2019clinically}.

Traditional medical image segmentation methods are mostly atlas-based. These methods usually rely on pre-computed templates, so they may not adequately account for the anatomical variance due to variations in organ shapes, removal of tissues, growth of tumor and differences in image acquisition. With the rise of deep learning, convolutional neural networks (CNNs) have been widely used in different domains of computer vision because of its extraordinary capability of extracting image features, such as object detection \cite{ren2016faster}, semantic segmentation \cite{fcn} and pose estimation \cite{cdg, gpa, ma2021transfusion}, etc. U-Net \cite{ronneberger2015unet} is the first to use CNNs in the field of medical image segmentation. Now U-Net and its variants \cite{resunet, unet++, unet3+} have achieved great success on this task.

Although CNNs are able to extract rich features, CNN-based approaches are not adequately equipped to encode long range interaction information \cite{tas}, whether within one single slice (intra-slice) or among the neighboring slices (inter-slice). In the field of medical image segmentation, it is useful to capture this information, since the texture, shape and size of many organs vary greatly across patients and it often requires long-range contextual information to reliably segment these organs. 

In the field of natural language processing (NLP), transformer-based methods \cite{vaswani2017attention} have achieved the state-of-the-art-performance in many tasks. Inspired by this design, researchers naturally think of leveraging Transformers' ability of modeling long range relationships to improve pure CNN-based models in natural images. However, less attention has been paid to use transformer-based models in medical image segmentation.

Recently, transformer-based models have been proposed in medical image segmentation in both 2D and 3D settings, with pros and cons associated with each as follows. In the 2D setting, TransUNet \cite{chen2021transunet} is the first to investigate the usage of Transformers for medical image segmentation to model long-range dependencies within a single 2D image. However, it does not consider the long-range dependencies in the 3D data, \ie, along the axial-axis, which is naturally provided in the 3D medical image data \cite{tang2021spatial}. In the 3D setting, CoTr \cite{xie2021cotr} is the first to explore Transformers to model long-range relationships in the volumetric data. However, because transformer modules and volumetric data both consume a lot of GPU memory, they need to compromise both in order to fit their model into easily accessible commodity GPUs. To address this, they cut the 3D volumetric data into local patches and process them one at a time, which results in loss of information from other patches. Moreover, they limit the pairwise attention to only a few voxels in the 3D data, which may be oversimplified and limit the ability of Transformers in modeling long-range relationships.

To better utilize Transformer to explore the long-range relationships in the 3D medical image data, in this paper, we propose \textbf{A}xial \textbf{F}usion \textbf{T}ransform\textbf{er} UNet (AFTer-UNet), an end-to-end medical image segmentation framework. Our motivation is to leverage both intra-slice and inter-slice contextual information to guide the final segmentation step. AFTer-UNet follows the U-shape structure of U-Net, which contains a 2D CNN encoder and a 2D CNN decoder. In between, we propose axial fusion transformer encoder to fuse contextual information in the neighboring slices. The axial fusion transformer encoder reduces the computational complexity by first separately calculating the attention along the axial axis and the attention within one single slice, and then fusing them together to produce the final segmentation map.

Our main contributions are listed as follows:

\vspace{-0.3em}
\begin{itemize}

    \item We propose an end-to-end framework, Axial Fusion Transformer UNet, to deal with 3D medical image segmentation tasks by fusing intra-slice and inter-slice information.
    \vspace{-0.2em}
    \item We introduce axial fusion mechanism, which reduces the computational complexity of calculating self-attention in 3D space.
    \vspace{-0.2em}
    \item We conduct extensive experiments on three multi-organ segmentation benchmarks, and demonstrate superior performance of AFTer-UNet compared to current transformer based models.
\end{itemize}


\section{Related work}
\vspace{-0.2em}
\subsection{CNN-based segmentation networks}
\vspace{-0.5em}
Early medical image segmentation methods are mainly contour-based and traditional machine learning-based algorithms. With the development of deep CNN, U-Net is proposed in \cite{ronneberger2015unet} for medical image segmentation. Due to the simplicity and superior performance of the U-shaped structure, various UNet-like methods are constantly emerging, such as Res-UNet \cite{resunet}, Dense-UNet \cite{denseunet}, U-Net++ \cite{unet++} and UNet3+ \cite{unet3+}. And it is also introduced into the field of 3D medical image segmentation, such as 3D U-Net \cite{3dunet} and V-Net \cite{milletari2016vnet}. It is also extended to other medical image analysis tasks, such as computer-aided diagnosis \cite{tang2018automated,tang2019nodulenet,tang2019end,liao2019evaluate,ding2017accurate}, image denoising \cite{you2019low,lyu2018super}, image registration \cite{balakrishnan2019voxelmorph,heinrich2019closing}, etc. At present, CNN-based methods have achieved tremendous success in the field of medical image segmentation due to its powerful representation ability \cite{nnunet,tang2021recurrent,gao2019focusnet,nikolov2021clinically,yancy9175642,xu2019lstm,you2020unsupervised,you2021momentum}. 

\subsection{Visual transformers}
\vspace{-0.5em}
Transformer was first proposed for the machine translation task in \cite{vaswani2017attention}. In the domain of natural language processing, the Transformer-based methods have achieved the state-of-the-art performance in various tasks \cite{devlin-etal-2019-bert,yancy2021form}. Motivated by the success of \cite{devlin-etal-2019-bert}, researchers introduced vision transformers (ViT) in \cite{dosovitskiy2020vit} for image classification tasks. Besides, \cite{timesformer} extended ViT to the field of video classification, which largely inspired our work. For object detection, \cite{detr} predicts the final set of detections by combining a common CNN with a transformer architecture. For semantic segmentation, \cite{SETR} exploited the transformer framework to implement the feature representation encoder by sequentializing images without using the traditional FCN \cite{fcn} design. 

\subsection{Transformers for medical image segmentation}
\vspace{-0.5em}
Recently, researchers have tried to apply transformer modules to improve the performance of current approaches.
TransUNet \cite{chen2021transunet} is the first paper to investigate the usage of Transformers for medical image segmentation problems. In this paper, the encoder and decoder of U-Net is connected by several Transformer layers. TransUNet leverages both CNN's capability of extracting low level features and Transformer's advantage of making high level sequence-to-sequence prediction. Swin-Unet \cite{cao2021swinunet} explores the application potential of pure transformer in medical image segmentation. However, both TransUNet and Swin-UNet only consider a single slice as input, so that the information along the axial-axis, which is intrinsically provided by a 3D volume, is not utilized. On the other hand, researchers have been trying applying transformers in a 3D way. However, computing self-attention directly on 3D space is not feasible due to the expensive computation. To resolve this issue, CoTr \cite{xie2021cotr} introduces deformable self-attention mechanism, which indeed reduces the computational complexity. However, the design of CoTr brings two issues. First, it requires 3D patches as inputs, which means a lot of information are lost due to the split of patches. Actually, this is a common issue for 3D medical image segmentation models. Second, CoTr computes self-attention over only $K$ locations. In their experiments, a larger $K$ leads to a higher Dice score. However, $K$ is only set to 4 as the maximum value. This is because the transformer module has to compromise more memory space for the expensive 3D convolutions. Therefore, the over-simplified way of computing self-attention in 3D space, may cost the loss contextual information. In our following experiments, CoTr shows marginal accuracy improvement compared to previous methods.


\section{Method}
\begin{figure*}
\begin{center}
\includegraphics[scale=0.5]{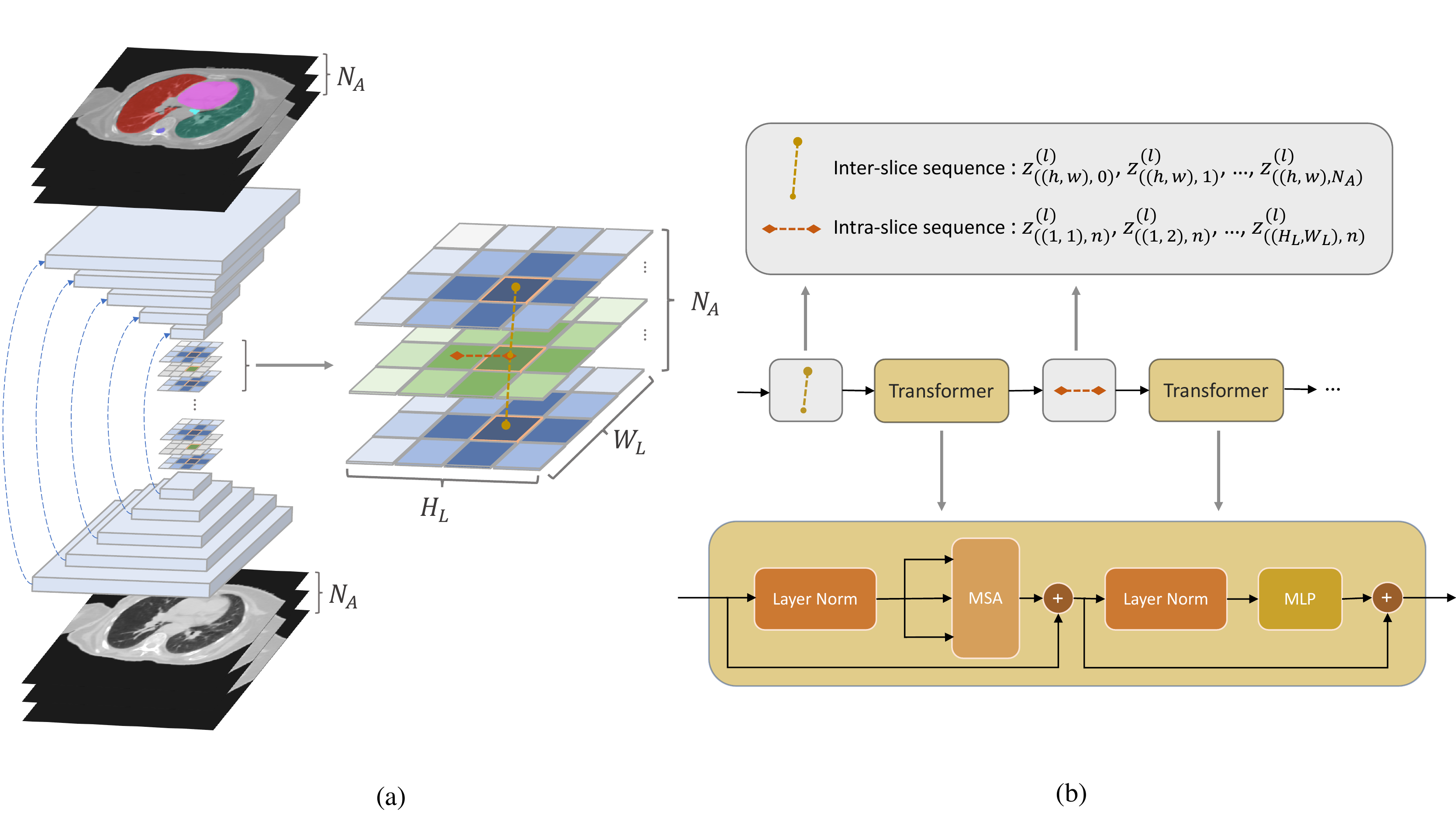}
\end{center}
\vspace{-1.5em}
\caption{Overview of Axial Fusion Transformer UNet. (a) The architecture of AFTer-UNet. We first encode the neighboring slice group $\mathbf{x}_d$ by the CNN encoder and get a corresponding feature map group $\mathbf{g}$. Then we apply axial fusion transformer to $\mathbf{g}$. Last, the feature group fused with both intra-slice and inter-slice cues are fed to the CNN decoder for segmentation. (b) The illustration of axial fusion mechanism. AFTer-UNet separately fuse the inter-slice and intra-slice information by Eq.\ref{eq:6} and Eq.\ref{eq:7} respectively. }
\label{fig:model}
\end{figure*}

\vspace{-0.5em}
Figure~\ref{fig:model} shows the details of AFTer-UNet. We follow the classic U-Net design, which includes a 2D CNN encoder for extracting fine-level image features and a 2D CNN decoder for achieving pixel-level segmentation. To better encode high-level semantic information, not only within a single slice but also among neighboring slices, we propose the axial fusion transformer in between. We now elaborate the details of each module in the following subsections.

\subsection{CNN encoder}
\vspace{-0.2em}
\subsubsection{Input formulation}
\vspace{-0.5em}
Given an input 3D CT scan $\mathbf{s} \in \mathbb{R}^{C \times H \times W \times D }$, we have a series of 2D slices along the axial axis, with height of $H$, width of $W$ and channel of $C=1$. The scan can be represented as $\mathbf{s} = \{\mathbf{s}_1, \mathbf{s}_2, ..., \mathbf{s}_D\}$, where $\mathbf{s}_d \in \mathbb{R}^{C \times H \times W}$. 
Firstly, for each 2D slice $\mathbf{s}_i$, we sample its $N_A$ neighboring slices along the axial axis by frequency $N_f$ and get $\mathbf{x} = \{\mathbf{x}_1, \mathbf{x}_2, ..., \mathbf{x}_D\}$, where $\mathbf{x}_d \in \mathbb{R}^{C \times H \times W \times N_A}$. 
For the each sampled neighboring slice group $\mathbf{x}_d$, we have $\mathbf{x}_d = \{\mathbf{s}_{a_0}, \mathbf{s}_{a_1}, ..., \mathbf{s}_{a_{N_A}}\}$, where $a_n = d - N_f \times (\frac{N_A}{2} - n)$ and $n \in \{0,1,...,N_A\}$. 

\vspace{-0.9em}
\subsubsection{Architecture}
\vspace{-0.5em}
The CNN encoder $\mathcal{E}^{CNN}$ mainly follows the design of U-Net, which includes $B$ blocks, connected by MaxPooling layers with both kernel size and stride of 2. Each block contains two Conv2d-ReLU pairs.
Additionally, we also add instance normalization layers in each pair between the 2d convolution layer and the ReLU layer. 
Given an input neighboring slice group $\mathbf{x}_d \in \mathbb{R}^{C \times H \times W \times N_A}$, each CNN encoder block $\mathcal{E}_{b}^{CNN}$ provides a corresponding feature map group $\mathbf{g}_{d,b} = \{\mathbf{g}_{d,b}^0, \mathbf{g}_{d,b}^1, ..., \mathbf{g}_{d,b}^{N_A} \} $, where $\mathbf{g}_{d,b}^{n}$ denotes a feature map  at level $b$ for slice $n$ and $\mathbf{g}_{d,b}^{n} \in \mathbb{R}^{C_b \times H_b \times W_b}$. Here we have $b \in \{1, 2, ..., B\}$ indicating the feature level, $H_b = \frac{H}{2^{b}}$,  $W_b = \frac{W}{2^{b}}$and $C_b$ denoting the height, width and number of channels at level $b$. 

However, here we only take the final feature map group $\mathbf{g}_{d,B} = \mathcal{E}^{C N N}(\mathbf{x}_{d})$ as input to the axial fusion transformer. 
We denote it as $\mathbf{g}$, where $\mathbf{g} \in \mathbb{R}^{C_L \times  H_L \times W_L \times N_A}$, for simplicity.
We choose this design for the following two reasons: First, taking the higher level feature map group means leveraging higher level semantic information, which is the motivation of applying transformers. Second, the GPU memory limits the size of the feature map group. 

\vspace{-0.2em}
\subsection{Axial fusion transformer encoder}
\vspace{-0.5em}
After extracting fine level features by the CNN encoder, we now introduce the Axial Fusion Transformer encoder to model high level semantic information not only within a single slice but also among neighboring slices along the axial axis.

\vspace{-1.1em}
\subsubsection{Feature maps as input embeddings}
\vspace{-0.5em}
In \cite{unetr}, not matter the input is a 2D slice/image or a 3D volume, it needs to be divided into small patches and then linearly mapped to vectors of a certain length. This is because transformers can't handle inputs with large size due to the memory limit. In our design, with the help of the above CNN encoder, we now have each feature map as an input with much smaller height $H_L$ and width $W_L$, so that each feature map $\mathbf{g}_n$ can be directly fed into the axial fusion transformer. Meanwhile, our approach provides more comprehensive information extracted from the whole image than from a single patch.

We then directly have $\mathbf{z}_{((h,w),n)}^{(0)}\in \mathbb{R}^{C_L}$ without the linear projection step in the original ViT \cite{dosovitskiy2020vit} setup: 
\begin{equation}\label{eq:1}
    \mathbf{z}_{((h,w),n)}^{(0)}= \mathbf{g}_{((h,w),n)}+\mathbf{e}_{((h,w), n)}^{p o s}
\end{equation}
, where $\mathbf{e}_{((h,w), n)}^{p o s} \in \mathbb{R}^{C_L}$ represents a learnable positional embedding to encode the vector location:(1) at $(h,w)$ within a single feature map  $\mathbf{g}_n$ and (2) at $n$ among feature maps in group $\mathbf{g}$.
The resulting sequence $\mathbf{z}_{((h,w),n)}^{(0)}$ for $(h,w) = (1,1), ..., (H_L, W_L)$ and $n = 0, 1, ..., N_A$ represents the input to the Transformer, and plays a role similar to the sequences of embedded words that are fed to text Transformers in NLP. Note that in our code implementation, the dimensions of height and width are flatten so the vector can also be represented as $\mathbf{z}_{(p,n)}^{(0)}$, where $p = W_L \cdot (h-1) + w$, but for illustration purpose, we keep the notion of $(h, w)$ here.

\vspace{-0.7em}
\subsubsection{Query-Key-Value matrices and self-attention}
\vspace{-0.5em}
The axial fusion transformer consists of $L$ blocks. For each block $l$, we compute each location's query/key/value vector from the representation $\mathbf{z}_{((h,w),n)}^{(l-1)}$, which is encoded by the preceding block:
\begin{equation}\label{eq:2}
    \mathbf{q}_{((h,w), n)}^{(l, a)}=W_{Q}^{(l, a)} \operatorname{LN}\left(\mathbf{z}_{((h,w), n)}^{(l-1)}\right) \in \mathbb{R}^{C_{h}}
\end{equation}
\begin{equation}\label{eq:3}
    \mathbf{k}_{((h,w), n)}^{(l, a)}=W_{K}^{(l, a)} \operatorname{LN}\left(\mathbf{z}_{((h,w), n)}^{(l-1)}\right) \in \mathbb{R}^{C_{h}}
\end{equation}
\begin{equation}\label{eq:4}
    \mathbf{v}_{((h,w), n)}^{(l, a)}=W_{V}^{(l, a)} \operatorname{LN}\left(\mathbf{z}_{((h,w), n)}^{(l-1)}\right) \in \mathbb{R}^{C_{h}}
\end{equation}
, where $LN()$ denotes LayerNorm \cite{ba2016layernorm}, $a \in \{1, 2, \ldots, \mathcal{A}\}$ represents the index over multiple attention heads and $\mathcal{A}$
denotes the total number of attention heads. Therefore, we have the dimensionality for each attention head as $C_{h} = C_L/\mathcal{A}$.

Self-attention weights are computed via dot-product. The self-attention weights $\boldsymbol{\alpha}_{((h,w), n)}^{(l, a)} \in \mathbb{R}^{(H_L \cdot W_L) \cdot N_A}$ for query at $((h,w), n)$ are:
\begin{equation}\label{eq:5}
    \boldsymbol{\alpha}_{((h,w), n)}^{(l, a)} = \operatorname{SoftMax}(\frac{\mathbf{q}_{((h,w), n)}^{(l, a)}}{\sqrt{C_{h}}}^{\top} \cdot\mathbf{k}_{\left((h,w)^{\prime}, n^{\prime}\right)}^{(l, a)})
\end{equation}
, where $(h,w)^{\prime} \in  \{(1,1), ..., (H_L, W_L)\}$ and $n^{\prime} \in \{ 0, 1, ..., N_A\}$. Note that when attention is computed only within a single feature map or only along the axial axis, the computation is significantly reduced. In the case of computing attention within a single feature map, only $H_L \cdot W_L$ query-key comparisons are made, using exclusively keys from the same feature map as the query:

\vspace{-0.5em}
\begin{equation}\label{eq:6}
    \boldsymbol{\alpha}_{((h,w), n)}^{(l, a) \textbf{intra}} = \operatorname{SoftMax}(\frac{\mathbf{q}_{((h,w), n)}^{(l, a)}}{\sqrt{C_{h}}}^{\top} \cdot\mathbf{k}_{\left((h,w)^{\prime}, n\right)}^{(l, a)})
\end{equation}
\vspace{-0.5em}
, where $(h,w)^{\prime} \in  \{(1,1), ..., (H_L, W_L)\}$. 

To get the encoding $\mathbf{z}_{((h,w),n)}^{(l)}$ at block $l$, we firstly compute the weighted sum of value vectors, using self-attention coefficients from each attention head:

\vspace{-0.5em}
\begin{equation}\label{eq:7}
        \mathbf{u}_{((h,w), n)}^{(l, a)}=\sum_{(h,w)^{\prime}=(1,1)}^{(H_L, W_L)} \sum_{n^{\prime}=1}^{N_A} \alpha_{((h,w), n),\left((h,w)^{\prime}, n^{\prime}\right)}^{(l, a)} \mathbf{v}_{\left((h,w)^{\prime}, n^{\prime}\right)}^{(l, a)}.
\end{equation}
These vectors from all heads are then concatenated, linearly projected by an fully connected layer ($\operatorname{FC}$) and passed through an multi-layer perceptron ($\operatorname{MLP}$) with layer norm ($\operatorname{LN}$). Residual connections are added after each operation:
\begin{equation}\label{eq:8}
    \mathbf{z}_{((h,w), n)}^{\prime(l)}=\operatorname{FC}\left(\begin{array}{c}
\mathbf{u}_{((h,w), n)}^{(l, 1)} \\
\vdots \\
\mathbf{u}_{((h,w), n)}^{(l, \mathcal{A})}
\end{array}\right)+\mathbf{z}_{((h,w), n)}^{(l-1)}
\end{equation}

\begin{equation}\label{eq:9}
    \mathbf{z}_{((h,w), n)}^{(l)}=\operatorname{MLP}\left(\operatorname{LN}\left(\mathbf{z}_{((h,w), n)}^{\prime(l)}\right)\right)+\mathbf{z}_{((h,w), n)}^{\prime(l)}
\end{equation}

\begin{table*}
\centering
\begin{tabular}{l|c|cccccc}
\toprule
Methods & DSC & Eso & Trachea & Spinal Cord & Lung(L) & Lung(R) & Heart\\
\midrule
U-Net\cite{ronneberger2015unet} & 91.18 & 78.85 & 90.72 & 89.37 & 97.31 & 96.37 & 94.46 \\
nnUNet-2D\cite{nnunet} & 89.74 & 78.82 & 88.32 & 86.61 & 96.03 & 96.65 & 92.01 \\ 
nnUNet-3D\cite{nnunet} & 91.63 & 81.18 & 89.32 & \bf{91.21} & 97.68 & 97.74 & 92.66 \\ 
Attention U-Net\cite{oktay2018attention} & 90.19 & 76.35 & 88.14 & 89.43 & 97.65 & 97.87 & 91.68\\
TransUNet\cite{chen2021transunet} & 91.38 & 78.27 & 91.45 & 88.36 & 97.63 & 97.84 & 94.74 \\  
Swin-Unet\cite{cao2021swinunet} & 91.26 & 78.98 & 91.20 & 88.64 & 97.64 & 97.79 & 93.30 \\  
CoTr\cite{xie2021cotr} & 91.39 & 79.06 & 91.55 & 88.67 & 97.47 & 97.65 & 93.92 \\  
\hline
AFTer-UNet & \bf{92.32} & \bf{81.47} & \bf{91.76} & 90.12 & \bf{97.80} & \bf{97.90} & \bf{94.86}\\ 
\bottomrule
\end{tabular}
\caption{Dice scores of different methods on in-house thorax-85 dataset.}
\label{table:thorax-85}
\end{table*}

\vspace{-1.2em}
\subsubsection{Fusing axial information}
\vspace{-0.5em}
Due to the limit of memory, computing self-attention over a 3D space by Eq.\ref{eq:5} is not feasible. Replacing it with 2D attention applied only on one single slice, \ie, Eq.\ref{eq:6} can certainly reduce the computational cost. However, such a model ignores to capture information among neighboring slices, which is naturally provided by a 3D volume. As shown in our experiments, considering less neighboring slices can provide poorer results. 

We propose axial fusion mechanism for computing attention along the axial axis, where the attention along the axial axis and the attention within a single slice are separately applied one after the other. By fusing the axial information this way, we firstly compute attention along the axial with all the channels at the same position at $(h,w)$: 
\vspace{-0.5em}
\begin{equation}
    \boldsymbol{\alpha}_{((h,w), n)}^{(l, a) \textbf{inter}} = \operatorname{SoftMax}(\frac{\mathbf{q}_{((h,w), n)}^{(l, a)}}{\sqrt{C_{h}}}^{\top} \cdot\mathbf{k}_{\left((h,w), n^{\prime}\right)}^{(l, a)})
\end{equation}
, where $n^{\prime} \in  \{1, ..., N_A\}$. The encoding $\mathbf{z}_{((h,w), n)}^{\prime(l)\textbf{inter}}$ resulting from the application of Eq.\ref{eq:8} using axial attention is then fed back for single slice attention computation instead of directly being passed to the MLP. In other words, new key/query/value vectors are obtained from $\mathbf{z}_{((h,w), n)}^{\prime(l)\textbf{inter}}$ and the single slice attention is then computed using Eq.\ref{eq:6}. Finally, the resulting vector $\mathbf{z}_{((h,w), n)}^{\prime(l)\textbf{intra}}$ is passed to the MLP of Eq.\ref{eq:9} to compute the final encoding $\mathbf{z}_{((h,w), n)}^{(l)}$ at position $((H,W), n)$ by block $l$. The final fused encoding for the feature map group $\mathbf{g}$ is $\mathbf{z}^{(L)}\in \mathbb{R}^{C \times H_L \times W_L \times N_A }$.

We learn distinct query/key/value matrices $\left\{W_{Q^{\text {slice }}}^{(l, a)}, W_{K^{\text {slice }}}^{(l, a)}, W_{V^{\text {slice }}}^{(l, a)}\right\}$ and $\left\{W_{Q^{\text {axial }}}^{(l, a)}, W_{K^{\text {axial }}}^{(l, a)}, W_{V^{\text {axial }}}^{(l, a)}\right\}$ over dimensions within one single slice and among slices along the axial axis. Note that compared to the $(H_L \cdot W_L) \cdot N_A$ comparisons each vector needed by the self-attention model of Eq.\ref{eq:5}, our approach performs only $(H_L \cdot W_L) + N_A$ comparisons per vector.

\vspace{-0.5em}
\subsection{CNN decoder}
\vspace{-0.3em}
The CNN decoder $\mathcal{D}^{CNN}$ of AFTer-UNet follows the design of U-Net as well, which is mostly symmetric to the CNN encoder $\mathcal{E}^{CNN}$. It includes $B$ Conv2d-ReLU blocks. Adjacent blocks are connected by Upsample layers with the scale factor of 2. Each block contains two Conv2d-ReLU pairs, with instance normalization layers between the 2d convolution layer and the ReLU layer. 

The final fused feature map group $\mathbf{z}^{(L)}$, provided by the last block of axial fusion transformer encoder, is taken as input to the CNN decoder. It gets progressively upsampled to the input resolution by the Upsample layers and gets refined by the Conv2d-ReLU blocks. As applied in the U-Net paper, we also the skip connections between encoder and decoder to keep more low-level details for better segmentation. 

Therefore, taking a series of sampled neighboring slice groups $\mathbf{x}$ as inputs, we now have $d$ segmentation map groups $\mathbf{y} = \{\mathbf{y}_1, \mathbf{y}_2,..., \mathbf{y}_D\}$ as outputs, where $\mathbf{y}_d \in \mathbb{R}^{C_{cls} \times H \times W \times N_A}$ and $C_{cls}$ denotes the number of organ classes. We only keep the middle segmentation map $\mathbf{y}_d^{\frac{N_A}{2}}$, remove its neighbor for all $d$ segmentation map groups and concatenate them together. At last, we have the final 3D prediction with respect to the 3D scan.

The loss function of our model is the sum of the dice loss and cross entropy loss.

\begin{table*}
\centering
\begin{tabular}{l|c|cccccccc}
\toprule
Methods & DSC & Aorta & Gallbladder & Kidney(L) & Kidney(R) & Liver & Pancreas & Spleen & Stomach\\
\midrule
U-Net\cite{ronneberger2015unet} & 74.68 & 87.74 & 63.66 & 80.60 & 78.19 & 93.74 & 56.90 & 85.87 & 74.16  \\ 

Attention U-Net\cite{oktay2018attention} & 75.57 & 55.92 & 63.91 & 79.20 & 72.71 & 93.56 & 49.37 & 87.19 & 74.95 \\

TransUNet\cite{chen2021transunet} & 77.48 & 87.23 & 63.13 & 81.87 & 77.02 & 94.08 & 55.86 & 85.08 & 75.62 \\

Swin-Unet\cite{cao2021swinunet} & 79.13 & 85.47 & \textbf{66.53} & 83.28 & 79.61 & \textbf{94.29} & 56.58 & 90.66 & \textbf{76.60} \\

CoTr\cite{xie2021cotr} & 78.46 & 87.06 & 63.65 & 82.64 & 78.69 & 94.06 & 57.86 & 87.95 & 75.74 \\
\hline
AFTer-UNet & \bf{81.02} & \bf{90.91} & 64.81 & \bf{87.90} & \bf{85.30} & 92.20 & \bf{63.54} & \bf{90.99} & 72.48\\
\bottomrule
\end{tabular}
\caption{Dice scores of different methods on the Synapse multi-organ CT (BCV) dataset.}
\label{table:bcv}
\end{table*}

\section{Experiments}
\vspace{-0.3em}
\subsection{Setup}
\vspace{-0.5em}
\textbf{Dataset} We conducted experiments using one abdomen CT dataset and two thorax CT datasets:

  - BCV is in the MICCAI 2015 Multi-Atlas Abdomen Labeling Challenge \cite{bcv}. It contains 30 3D abdominal CT scans from patients with various pathologies and has variations in intensity distributions between scans. Following \cite{chen2021transunet, cao2021swinunet}, we report the average DSC on 8 abdominal organs (aorta, gallbladder, spleen, left kidney, right kidney, liver, pancreas, spleen, stomach) with a random split of 18 training cases and 12 test cases.
  
  - Thorax-85 is an in-house dataset from \cite{thorax-dataset} that contains 85 3D thorax CT images. We report the average DSC on 6 thorax organs (eso, trachea, spinal cord, left lung, right lung,  and heart) with a random split of 60 training cases and 25 test cases.
  
  - SegTHOR is from the 2019 Challenge on Segmentation of THoracic Organs at Risk in CT Images \cite{lambert2019segthor}. It contains 40 3D thorax CT scans. We report the average DSC on 4 thorax organs (eso, trachea, aorta, and heart) with a random split of 30 training cases and 10 validation cases.

\textbf{Evaluation metric} We use the same evaluation metric Sørensen–Dice coefficient (DSC) as in previous work \cite{chen2021transunet, xie2021cotr}. DSC measures the overlap of the prediction mask $\mathbf{m}_p$ and ground truth mask $\mathbf{m}_g$ and is defined as:
\vspace{-0.9em}
\begin{equation}\label{dsc}
    \operatorname{DSC}(\mathbf{m}_p, \mathbf{m}_g)=\frac{2|\mathbf{m}_p \cup \mathbf{m}_g|}{|\mathbf{m}_p|+|\mathbf{m}_g|}
\end{equation}
\vspace{-1em}

\textbf{Implementation details} All images are resampled to have spacing of 2.5mm × 1.0mm × 1.0mm, with respect to the depth, height, and width of the 3D volume.
In the training stage, we apply elastic transform for alleviating overfitting. We use Adam\cite{adam} optimizer with momentum of $0.9$ and weight decay of $10^{-4}$ to train AFTer-UNet for 550 epochs. The learning rate is set to $10^{-4}$ for the first 500 epochs and $10^{-5}$ for the last 50 epochs. In one epoch, for each 3D CT scan $s$, we only randomly select one slice group $\mathbf{x}_d$, rather than all of them.  We set the number of Conv2d-ReLU blocks  $B = 5$, number of axial fusion transformer $L=6$, number of attention heads $\mathcal{A} = 8$, number of neighboring slices $N_A = 8$ and sample frequency $N_f = 1$.

\begin{table}
\centering
\begin{tabular}{l|c|cccc}
\toprule
Methods & DSC & Eso & Trachea & Aorta & Heart\\
\midrule
U-Net & 89.97 & 80.07 & 91.23 & 94.73 & 93.83 \\

Att U-Net & 90.47 &  81.25 & 90.82 & 94.74 & 95.07 \\
TransUNet & 91.50 &  81.41 & 94.05 & 94.48 & 96.07 \\  
Swin-Unet & 91.29 &  81.06 & 93.27 & 94.82 & 96.02 \\  
CoTr & 91.41 &  81.53 & 94.03 & 94.06 & 96.01 \\  
\hline
AFTer-UNet & \bf{92.10} & \bf{82.98} & \bf{94.20} & \bf{94.92} & \bf{96.31} \\  
\bottomrule
\end{tabular}
\caption{Dice scores of different methods on SegTHOR thorax dataset.}
\label{table:segthor}
\end{table}

\begin{table*}[hbt!]
\centering
\begin{tabular}{c|c|cccccc}
\toprule
$N_A$ & DSC & Eso & Trachea & Spinal Cord & Lung(L) & Lung(R) & Heart\\
\midrule
1 & 91.44 & 78.31 & 90.65 & 90.14 & 97.48 & 97.69 & 94.35\\
2 & 91.66 & 78.54 & 91.35 & 90.27 & 97.59 & 97.60 & 94.59\\ 
4 & 91.98 & 79.74 & 91.42 & 90.71 & 97.59 & 97.77 & 94.66\\ 
8 & \bf{92.32} & 81.47 & 91.76 & 90.12 & 97.80 & 97.90 & 94.86\\  
\bottomrule
\end{tabular}
\vspace{-0.5em}
\caption{Ablation study on $N_A$, the number of neighboring axial slices.}
\label{table:N_A}
\end{table*}

\begin{table*}[hbt!]
\centering
\begin{tabular}{c|c|cccccc}
\toprule
$L$ & DSC & Eso & Trachea & Spinal Cord & Lung(L) & Lung(R) & Heart\\
\midrule
1 & 91.19 & 80.47 & 91.38 & 87.6 & 96.43 & 96.38 & 94.89\\
2 & 92.13 & 80.54 & 91.40 & 90.64 & 97.63 & 97.76 & 94.82\\
4 & 92.25 & 80.66 & 91.66 & 90.78 & 97.70 & 97.83 & 94.88\\
6 & \bf{92.32} & 81.47 & 91.76 & 90.12 & 97.80 & 97.90 & 94.86\\  
\bottomrule
\end{tabular}
\vspace{-0.5em}
\caption{Ablation Study on $L$, the number of transformer layers.}
\label{table:L}
\end{table*}

\begin{table*}[hbt!]
\centering
\begin{tabular}{c|c|cccccc}
\toprule
$N_f$ & DSC & Eso & Trachea & Spinal Cord & Lung(L) & Lung(R) & Heart\\
\midrule
1 & \bf{92.32} & 81.47 & 91.76 & 90.12 & 97.80 & 97.90 & 94.86\\  
2 & 91.92 & 79.77 & 91.06 & 90.49 & 97.69 & 97.65 & 94.87\\
4 & 92.05 & 79.72 & 91.69 & 90.30 & 97.81 & 97.88 & 94.89\\
\bottomrule
\end{tabular}
\vspace{-0.5em}
\caption{Ablation study on $N_f$, the sampling frequency on axial axis.}
\label{table:N_f}
\end{table*}

\vspace{-0.5em}
\subsection{Results on Thorax-85}

Table \ref{table:thorax-85} shows the performance comparison of AFTer-UNet with previous work on Thorax-85. We ran the following representative algorithms: U-Net \cite{ronneberger2015unet}, Attention U-Net \cite{oktay2018attention}, nnU-Net \cite{nnunet}, TransUNet \cite{chen2021transunet}, Swin-Unet \cite{cao2021swinunet}, and CoTr \cite{xie2021cotr}. U-Net is a well-established medical image segmentation baseline algorithm. Attention U-Net \cite{oktay2018attention} is a multi-organ segmentation framework that uses gated attention to filter out irrelevant responses in the feature maps. nnU-Net \cite{nnunet} is a self-adaptive medical image semantic segmentation framework that wins the first in the Medical Segmentation Decathlon(MSD) challenge \cite{simpson2019large}. TransUNet \cite{chen2021transunet} presents the first study which explores the potential of transformers in the context of 2D medical image segmentation. Swin-Unet \cite{cao2021swinunet} explores using pure transformer modules on 2D medical image segmentation tasks, without any convolutional layers. CoTr \cite{xie2021cotr} firstly explores transformer modules for 3D medical image segmentation. The above-mentioned works cover a wide range of algorithms for multi-organ segmentation and should provide a comprehensive and fair comparison to our method on the in-house Thorax-85 dataset.

By comparing the results on left lung, right lung and heart, all models provide comparable results. This is because those organs are usually large and have regular shapes. However, for organs like esophagus and trachea, 3D models and transformer-based models have consistently higher DSC. These organs often have more anatomical variance, so the capability of long-range sequence modeling provides a holistic understanding of the context, which is beneficial. Both CoTr and AFTer-UNet consider using transformers to fuse 3D information. However, CoTr directly takes 3D patches as inputs, which may cause the loss of spatial information inter-patches. Besides, due to CoTr's heavy 3D convolution operation, they limit the pairwise attention to only a few voxels, which may be oversimplified and limit the ability of Transformers in modeling long-range relationships. AFTer-UNet, however, applies 2D convolution to extract fine-level detail features and leave more memory space for the axial fusion transformer to extract richer inter-slice and intra-slice information. On our in-house thorax-85 dataset, the higher capability of long dependency modeling enables AFTer-UNet to outperform the previous state-of-the-art transformer-based method CoTr by 0.95\%. Altogether, we demonstrated the effectiveness of the proposed method, which achieves an average DSC of 92.32\% on six thorax organs.

\vspace{-0.5em}
\subsection{Results on public datasets}
\vspace{-0.5em}
We also conduct experiments on a public abdomen dataset, BCV, and a public thorax dataset, SegTHOR. Table \ref{table:bcv} and \ref{table:segthor} show the performance of AFTer-UNet and previous models. For large and normal-shaped organs, such as liver, spleen, stomach, and heart, all models are on par with each other. This is consistent with the conclusion we draw in section 4.2. However, for organs like aorta, left kidney, right kidney and pancreas, AFTer-UNet outperforms U-Net baseline by 3.17\%, 7.30\%, 7.11\%, and 6.64\% respectively and outperforms CoTr by 3.85\%, 5.26\%, 6.61\% and 5.68\% respectively. On average, AFTer-UNet has 4.34\% improvement compared to U-Net and 2.56\% improvement compared to CoTr. For elongated shaped organs like the esophagus in SegTHOR, AFTer-UNet provides 2.91\% improvement compared to U-Net baseline and 1.45\% improvement compared to CoTr. On average, our AFTer-UNet outperforms U-Net by 2.13\% and CoTr by 0.69\%.

\subsection{Ablation study on Thorax-85}
\vspace{-0.5em}
We further conduct extensive ablation studies on Thorax-85 to explore the influence of different hyperparameters:

\textbf{The number of neighboring axial slices $N_A$.} 
As discussed in previous sections, AFTer-UNet fuses inter-slice information by using the axial fusion mechanism. The number of neighboring axial slices is an essential factor of the mechanism. It is observed from Table \ref{table:N_A} that the more neighboring slices we fuse, the higher the dice score is. Note that for elongated shaped organs such as esophagus and trachea, the dice scores increase more obviously than other large and normal shaped organs such as left lung, right lung and heart. It makes sense because those organs with large anatomical variances require more global information and increasing the number of neighboring slices can help to fulfill this requirement. In our AFTer-UNet model, we use $N_A = 8$ as the number of neighboring axial slices.

\textbf{The number of axial fusion transformer layers $L$.}
Table \ref{table:L} shows the results with $L = 1, 2, 4, 6$. It is observed that the average dice scores are improved when the number of axial fusion transformer layers goes up. Especially for elongated shaped organs such as esophagus and trachea, as $L$ increases, dice scores on these two organs are improved more obviously than other large organs with normal shapes. This again shows the effectiveness of the axial fusion mechanism on compounding inter-slice cues.

\textbf{The sampling frequency on axial axis $N_f$.} 
We conduct experiments with various $N_f = 1, 2, 4$, and results are shown in Table \ref{table:N_f}. It turns out that increasing the sampling frequency will heart AFTer-UNet's performance. In our design of axial fusion mechanism, lower $N_f$ leads to \textbf{denser} inter-slice information. The results show that it's more important to fuse the nearest neighboring information than slices far away. This might be one of the reasons why AFTer-UNet outperforms CoTr. The latter only considers several key points information, which are \textbf{sparsely} distributed in a 3D volume.

\begin{figure*}[hbt!]
\centering
\includegraphics[scale=0.71]{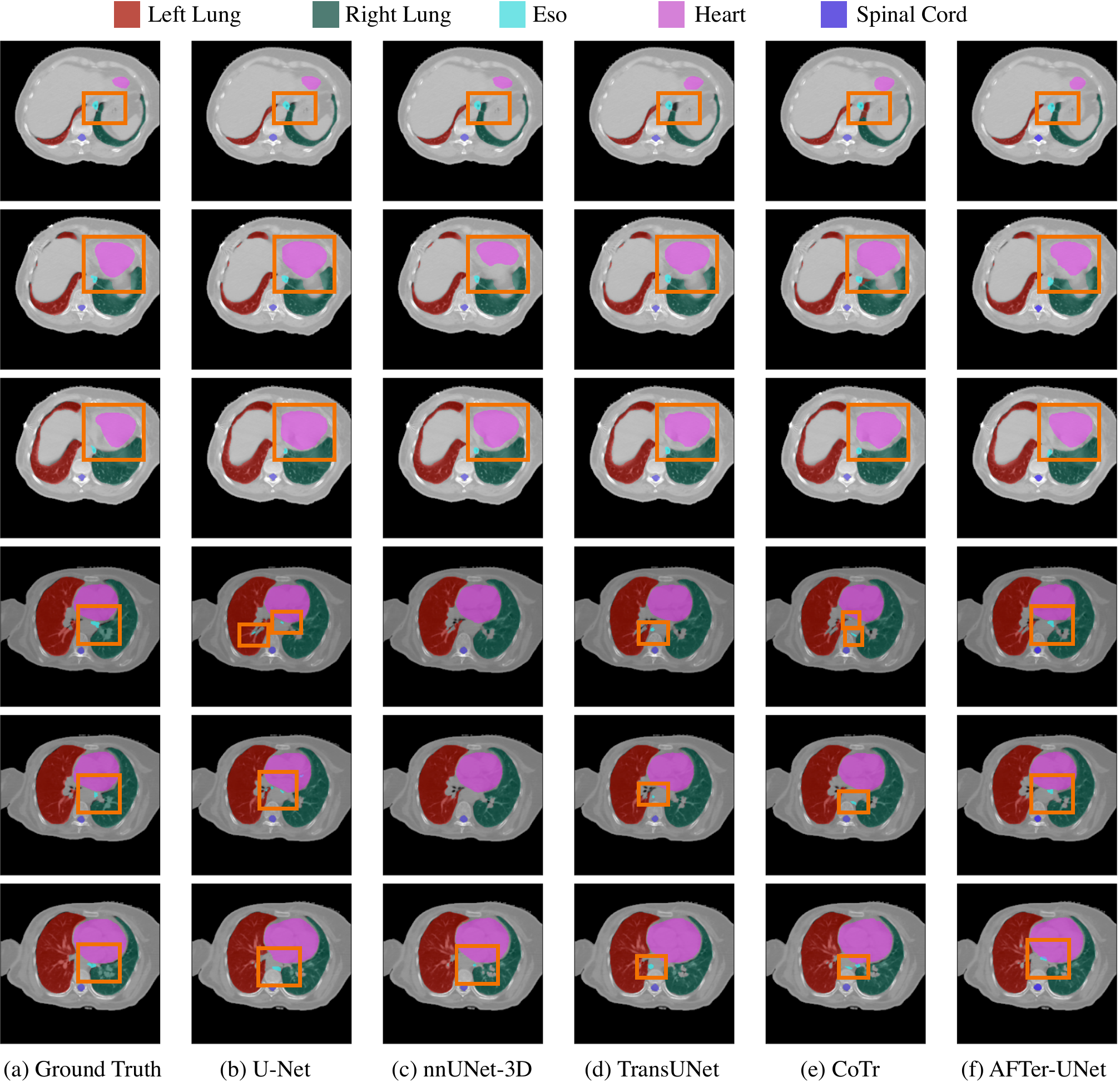}
\vspace{-0.5em}
\caption{Qualitative results of different approaches on Thorax-85 dataset. (a) shows the ground truth of the CT slice. (b)-(e) show the results of previous methods. (f) shows the results of AFTer-UNet. The regions in orange rectangles indicate the effectiveness to our model.}
\label{fig:qualitative}
\end{figure*}

\begin{figure}[hbt!]
    \centering
    \includegraphics[scale=0.41]{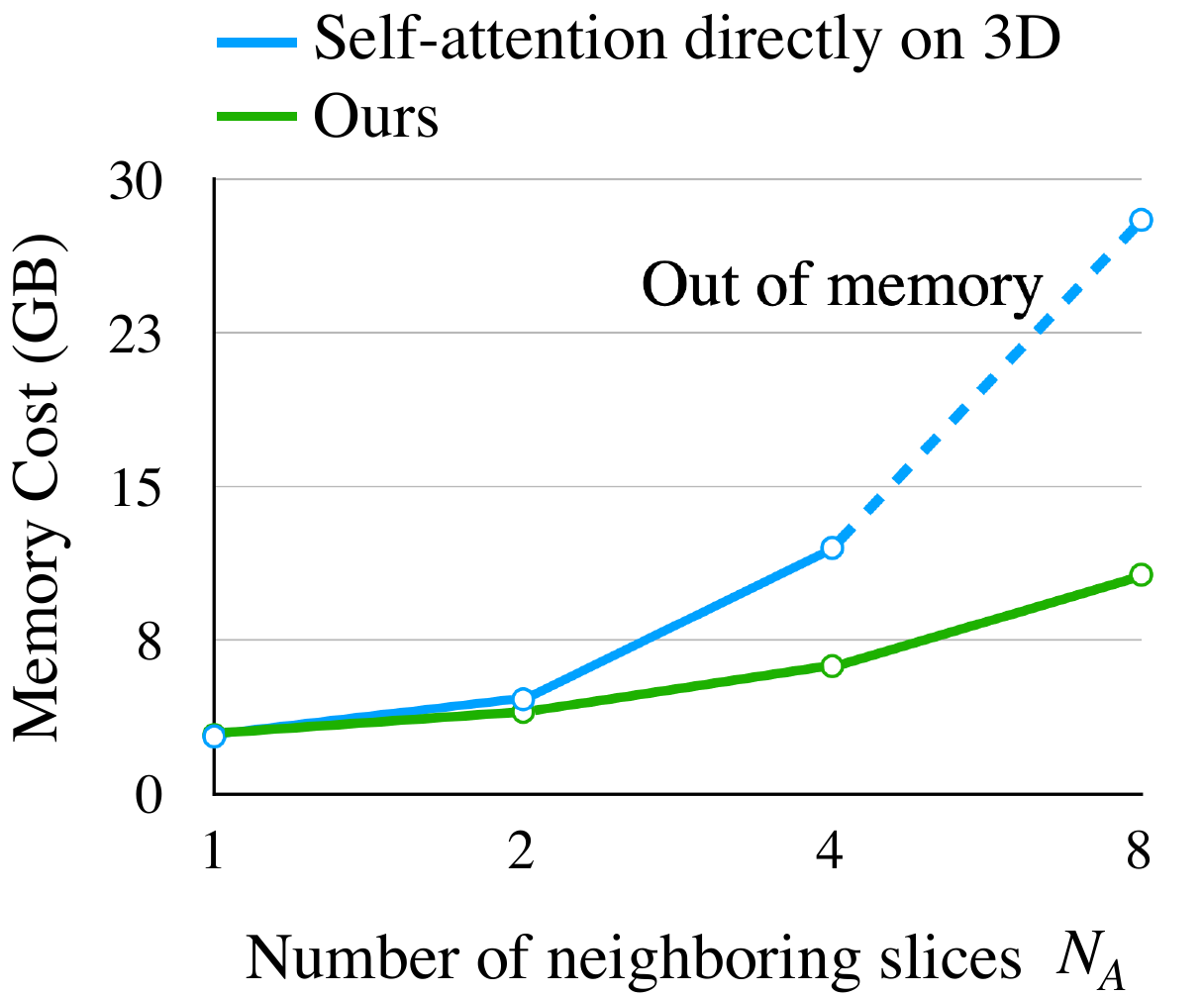}
    \caption{GPU memory consumption comparison between computing self-attention on 3D directly and computing by our proposed axial fusion mechanism.}
    \label{fig:mem_cost}
\end{figure}

\vspace{-0.3em}
\subsection{Qualitative results}
\vspace{-0.5em}
Fig.\ref{fig:qualitative} shows the qualitative results of different approaches on Thorax-85 dataset. Thanks to the axial fusion mechanism, AFTer-UNet presents its effectiveness compared to other methods. We may focus on esophagus,  the hardest to segment in Thorax-85, due to its large anatomy variance and elongated shape. For previous methods, segmentation maps might move largely between slices, which is unreasonable. However, AFTer-UNet (last column) provides consecutive and accurate predictions by considering the inter-slice and intra-slice context. Note that in this part, we didn't visualize trachea since 1) trachea is relatively naive to segment so all model provides accurate results and 2) trachea lies in the very top region of thorax where few other organs can be shown at the same time.

\vspace{-0.3em}
\subsection{Memory consumption and model parameters}
\vspace{-0.5em}
We compare the GPU memory consumption of 1) directly computing self-attention on 3D and 2) computing by our proposed axial fusion mechanism. As shown in Fig.\ref{fig:mem_cost}, our proposed axial fusion mechanism leads to dramatic computational savings. Therefore, our AFTer-UNet is able to be trained on a single RTX-2080Ti GPU with 11GB memory. Besides, TransUNet has 43.5M parameters and CoTr has 41.9M parameters \cite{xie2021cotr}. Meanwhile, AFTer-UNet has 41.5M parameters. This shows our method doesn't include more parameters to achieve its effectiveness than previous transformer based models.

\vspace{-0.4em}
\section{Conclusion}
\vspace{-0.5em}
In this work, we introduce AFTer-UNet, an end-to-end framework for medical image segmentation. The proposed framework use an axial fusion mechanism to fuse intra-slice and inter-slice contextual information and guide the final segmentation process. Experiments on three datasets demonstrate our model's effectiveness compared to previous work.

{\small
\bibliographystyle{ieee_fullname}
\bibliography{egbib}
}

\end{document}